\documentclass{elsart}
\usepackage{graphicx}

\begin{document}

\begin{frontmatter}

\title{\bf Event-by-event $p_T$ fluctuations and multiparticle clusters in
relativistic heavy-ion collisions\thanksref{grant}} 
\thanks[grant]{Research supported in part by: the Polish State 
Committee for Scientific Research, grant No. 2 P03B 059 25; Funda\c{c}\~{a}o para
a Ci\^{e}nca e a Tecnologia, POCTI/FNU/50336/2003, POCI/FP/63412/2005; PRAXIS
XXI/BCC/429/94, and GRICES}
\author[as,ifj]{Wojciech Broniowski,} 
\author[uc]{Brigitte Hiller,}
\author[as,ifj]{Wojciech Florkowski,} 
\author[ifj]{and Piotr Bo\.zek}
\address[as]{Institute of Physics, \'Swi\c{e}tokrzyska Academy,
ul.~\'Swi\c{e}tokrzyska 15, PL-25406~Kielce, Poland} 
\address[ifj]{The H. Niewodnicza\'nski Institute of Nuclear Physics, 
Polish Academy of Sciences, PL-31342 Krak\'ow, Poland}
\address[uc]{Centro de F{\'{i}}sica Te\'orica, Departamento de F{\'{i}}sica, 
University of Coimbra, P-3004-516 Portugal}

\begin{abstract} 
We explore the dependence of the $p_T$ correlations in the event-by-event analysis of 
relativistic heavy-ion collisions at RHIC made recently by the PHENIX and STAR 
Collaborations. We point out that the observed scaling of strength of dynamical
fluctuations with the inverse number of particles can be naturally 
explained by the formation of clusters. We argue that the large magnitude of 
the measured covariance implies that the clusters contain at least several particles. We 
also  discuss whether the clusters may originate from jets. In addition, we  provide numerical estimates 
of correlations coming from resonance decays and thermal clusters. 

{\it Keywords: } relativistic heavy-ion collisions,
event-by-event fluctuations, 
particle correlations, 
\end{abstract}

\end{frontmatter}

\vspace{-7mm} PACS: 25.75.-q, 25.75.Gz, 24.60.-k


\maketitle

Recently new data on the event-by-event fluctuations have been provided by the 
PHENIX \cite{Adcox:2002pa} and STAR \cite{Adams:2003uw,Adams:2005ka} Collaborations, 
shedding more light on the previously accumulated knowledge in the field 
\cite{Gazdzicki:1992ri,Stodolsky:1995ds,Shuryak:1997yj,%
Mrowczynski:1997kz,Stephanov:1999zu,Voloshin:1999yf,%
Appelshauser:1999ft,Korus:2001au,%
Baym:1999up,Bialas:1999tv,Asakawa:2000wh,Heiselberg:2000fk,%
Pruneau:2002yf,Jeon:2003gk,Gavin:2003cb,Abdel-Aziz:2005wc}. 
One of the most fascinating but intricate questions is whether the $p_T$
fluctuations in large windows of pseudorapidity and azimuthal angle
at intermediate momenta can result from jets \cite{Adler:2003xq,Liu:2003jf}. 
In this letter we explore  the basic 
facts of the recent data \cite{Adcox:2002pa,Adams:2005ka}. In particular, we argue that 
since $i$)~the mean and the variance of the inclusive momentum distribution are 
practically constant at low centrality parameters, then $ii$)~the variance of the 
average momenta for the mixed events is practically equal to the variance of the 
inclusive distribution divided by the average multiplicity. Moreover, and this is our 
basic observation, $i$) also results in the fact that $iii$)~the difference of the 
experimental and mixed-event variances of average $p_T$, denoted as 
$\sigma_{\rm dyn}^2$, scales as inverse multiplicity, as seen in experiments 
\cite{Adcox:2002pa,Adams:2005ka}. 
A possible explanation of this scaling can be provided by clustering in 
the expansion velocity: matter expands in ``lumped clusters'' of chunks of matter, 
having close collective velocity within a cluster, which induces correlations. Moreover, 
we show that the value of $\sigma_{\rm dyn}$ is {\em large} at the expected scale provided by 
the variance of $p_T$, which indicates that the clusters should contain at least several 
particles in order to combinatorically enhance the magnitude to the observed level. 
We discuss whether jets may be responsible for the formation of the clusters. 
Finally, we compute numerically the value of $\sigma_{\rm dyn}^2$ coming from the 
resonance decays and from thermal clusters in statistical models of heavy-ion collisions. 
The found values of the covariance per pair are small, suggesting larger numbers of 
particles in clusters.

We begin by exploring the PHENIX measurement \cite{Adcox:2002pa} of the event-by-event 
fluctuations of the transverse momentum at $\sqrt{s_{NN}}=130$~GeV. 
To simplify our notation, the letter $p$ is used to denote 
$|\vec{p}_T|$, $p_i$ is the value of $p$ for the $i$th particle, and $M = \sum_{i=1}^n p_i/n $ is 
the average transverse momentum in an event of multiplicity $n$. The PHENIX results are recalled
in Table \ref{tab:data}. Several features of the data are striking: the 
quantities $\langle M \rangle$ and $\sigma_p$ are practically constant in the 
reported centrality range $c=0-30$\%,
\begin{eqnarray}
\langle M \rangle = \hbox{const.}\,, \;\;\;\;\;  \sigma_p = \hbox{const.}
~~(\rm{at~low}~c).  \label{const}
\end{eqnarray}
We call the range of $c$ where (\ref{const}) holds the ``fiducial centrality range'' - 
this is where our conclusions will be drawn. We note 
that  for peripheral events  incomplete thermalization can result in a 
different strength of $p_T$ fluctuations
\cite{Abdel-Aziz:2005wc}.
 Next, we observe that 
$\sigma_M \simeq \sigma_p/\sqrt{\langle n \rangle}$. More precisely, for 
the mixed events one finds the formula
\begin{eqnarray}
\sigma_M^{\rm mix} \simeq \sigma_p \sqrt{\frac{1}{  \langle n \rangle }
+\frac{\sigma_n^2}{\langle n \rangle^3}}, \label{varmix}
\end{eqnarray}
working at the level of $1-2$\%.
Finally, the difference of the experimental and mixed-event variances of 
average $p_T$, denoted as $\sigma_{\rm dyn}^2$, scales 
to a remarkable accuracy as the inverse multiplicity,
\begin{eqnarray}
\sigma_{\rm dyn}^2 \equiv \sigma_M^2 - \sigma_M^{2, {\rm mix}} 
\sim \frac{1}{\langle n \rangle}. 
\label{scaledyn} 
\end{eqnarray}

\begin{table}[b]
\caption{\label{tab:data} 
Analysis  of the event-by-event fluctuations in the transverse momentum. Upper rows: 
the PHENIX experimental data at $\sqrt{s_{NN}} =$ 130~GeV  \cite{Adcox:2002pa}; 
middle rows: the mixed-event results; bottom rows: our way of looking at the data. 
One observes that to a good approximation
$ \sigma_M^{2, \rm mix} \simeq \sigma_p^2/\langle n \rangle$ and 
$ \sigma_{\rm dyn}^2 = (\sigma_M^2-\sigma_M^{2,{\rm mix}}) \sim 1/\langle n \rangle$. 
Except for the first two rows, all values are given in MeV. The errors in the last row 
reflect the unknown round-off errors in the data of the upper and middle parts.}
\begin{center}
\begin{tabular}{|l|c|c|c|c|}
\hline
centrality & 0-5\% & 0-10\% & 10-20\% & 20-30\% \\
\hline
$\langle n \rangle$ & 59.6 & 53.9 & 36.6 & 25.0 \\
$\sigma_n$ & 10.8 & 12.2 & 10.2 & 7.8 \\
$\langle M \rangle$ & 523 & 523 & 523 & 520 \\
$\sigma_p$ & 290 & 290 & 290 & 289 \\
$\sigma_M$ & 38.6 & 41.1 & 49.8 & 61.1 \\
\hline
$\langle M \rangle^{\rm mix}$ & 523 & 523 & 523 & 520 \\
$\sigma_M^{\rm mix}$ & 37.8 & 40.3 & 48.8 & 60.0 \\
\hline
$\sigma_p \sqrt{\frac{1}{\langle n \rangle}+\frac{\sigma_n^2}{\langle n \rangle^3}}$ & 38.2 & 40.5 & 49.8 & 60.8 \\
$\sigma_{\rm dyn} \sqrt{ \langle n \rangle}$ & $60.3 \pm 1.6$ & $59.2 \pm 1.5$ & 
$59.8 \pm 1.2$ & $57.7 \pm 1.1$ \\
\hline
\end{tabular}
\end{center}
\end{table}

Now we proceed to elementary statistical considerations. Consider events of multiplicity 
(of charged particles) $n$ 
and transverse momenta $p_1, p_2, \dots, p_n$. The multiplicity $n$ and the momenta are varying 
randomly from event to event.  The probability density of occurrence of a given momentum
configuration is $P(n) \rho_n(p_1,\dots,p_n)$, where $P(n)$ is the 
multiplicity distribution
and $\rho_n(p_1,\dots,p_n)$ is the conditional 
probability distribution of occurrence of $p_1,\dots,p_n$ in accepted events, 
provided we have the multiplicity $n$. Note that in general $\rho$ depends functionally on $n$, 
which is indicated by the subscript. The normalization is
\begin{equation}
\label{norm}
\sum_{n} P(n)=1, \;\;\;\;
\int dp_1 \dots dp_n \rho_n(p_1,\dots,p_n)=1.
\end{equation}
The {\em marginal} probability densities are defined as
\begin{eqnarray}
\rho_n^{(n-k)} (p_1,\dots,p_{n-k}) \equiv \int dp_{n-k+1} \dots dp_n \rho_n(p_1,\dots,p_n),
\end{eqnarray} 
with $k=1,\dots,n-1$. These are also normalized to unity, as follows from Eq.~(\ref{norm}).
Since the number of arguments distinguishes the marginal distributions $\rho_n^{(n-k)}$,
in the following we drop the superscript $(n-k)$.  Further, we introduce the
following definitions
\begin{eqnarray}
\langle p \rangle_n &\equiv& \int dp \rho_n(p) p, \quad
{\rm var}_n(p) \equiv \int dp \rho_n(p) \left (p-\langle p \rangle_n \right )^2 , \nonumber \\
{\rm cov}_n(p_1,p_2) &\equiv& \int dp_1 dp_2 \left (p_1-\langle p \rangle_n \right ) 
\left (p_2-\langle p \rangle_n \right)\rho_n(p_1,p_2). 
\end{eqnarray}
The subscript $n$ indicates that the averaging is taken in samples of multiplicity $n$. 
We note in passing that the commonly used {\em inclusive} distributions are related 
to the marginal probability distributions in the following way:
\begin{eqnarray}
\hspace{-0.8cm} \rho_{\rm in}(x) &\equiv&  
\sum_n P(n) \int dp_1 \dots dp_n \sum_{i=1}^n \delta(x-p_i) 
\rho_n(p_1,\dots,p_n) 
=  \sum_n n P(n) \rho_n(x), \nonumber \\
\hspace{-0.8cm} \rho_{\rm in}(x,y) &\equiv& 
\sum_n P(n) \int dp_1 \dots dp_n \sum_{i,j=1, j \neq i}^n 
\delta(x-p_i) \delta(y-p_j) \rho_n(p_1,\dots,p_n) \nonumber \\
&=& \sum_n n(n-1) P(n) \rho_n(x,y),
\label{incl}
\end{eqnarray}
which are normalized to $\langle n \rangle$ and $\langle n(n-1)\rangle$, respectively. 

For the variable $M= \sum_{i=1}^n p_i/n $ we find immediately
%
\begin{eqnarray}
\langle M \rangle &=& \sum_n P(n)\int dp_1 \dots dp_n M \rho_n(p_1,\dots,p_n) 
=\sum_n P(n) \langle p \rangle_n, \nonumber \\
\langle M^2 \rangle &=& \sum_n P(n)\int dp_1 \dots dp_n M^2 \rho_n(p_1,\dots,p_n) 
\\
&=&
\sum_n \frac{P(n)}{n} \langle p^2 \rangle_n  + 
\sum_n \frac{P(n)}{n^2} \left [ \sum_{i,j=1, j \neq i}^n {\rm cov}_n(p_i,p_j) + 
n(n-1)\langle p \rangle_n^2 \right ]. \nonumber
\end{eqnarray}

\begin{figure}[b]
\begin{center}
\includegraphics[width=7.5cm]{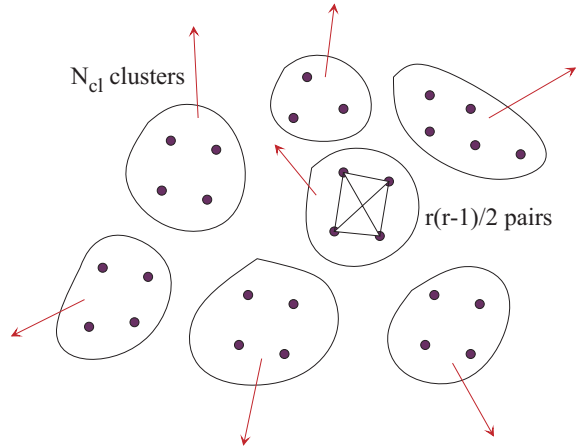}
\end{center}
\caption{The cluster model of correlations. Particles are grouped in $N_{\rm cl}$ clusters, 
containing on the average $r$ particles. The particles within a cluster move at very similar 
collective velocities, indicated by arrows.}
\label{fig:clust}
\end{figure}

Next, we use the experimental fact that the variance of the momentum distribution and its mean
are independent of centrality in the fiducial range, 
which allows us to replace the quantities $\langle p \rangle_n$ by $\langle M \rangle$ and 
$\langle p^2 \rangle_n - \langle p \rangle_n^2$ by $\sigma_p^2$ at the
average multiplicity, denoted
as $\sigma^2_{p, \langle n \rangle}$. In this way we get
\begin{eqnarray}
\sigma_M^2 = \sigma_{p, \langle n \rangle}^2 \sum_n \frac{P(n)}{n} + 
\sum_n \frac{P(n)}{n^2} \left [ \sum_{i,j=1, j \neq i}^n   {\rm cov}_{n}(p_i,p_j) 
\right ]. 
\label{central}
\end{eqnarray}
In the mixed events, by construction, particles are not correlated, hence the covariance term 
in Eq.~(\ref{central}) vanishes and
\begin{eqnarray}
\sigma_M^{2,{\rm mix}} = \sigma_{p, \langle n \rangle}^2 \sum_n \frac{P(n)}{n} \simeq 
\sigma_{p, \langle n \rangle}^2 \left ( \frac{1}{ \langle n \rangle } 
+ \frac{\sigma_n^2}{\langle n \rangle^3} +
\dots \right ), 
\label{mixed}
\end{eqnarray}
where in the last equality we have used the fact that the distribution $P(n)$ 
is narrow and expanded
$1/n=1/[\langle n \rangle +(n-\langle n \rangle)]$ to second order in $(n-\langle n \rangle)$. Comparison made in
Table \ref{tab:data} shows that formula (\ref{mixed}) works at the 1-2\% level. In addition,
since $\sigma_{p, \langle n \rangle}$ is not altered by the event mixing
procedure, subtracting (\ref{mixed}) from (\ref{central}) yields
\begin{eqnarray}
\sigma_{\rm dyn}^2 =
\sum_n \frac{P(n)}{n^2}  \sum_{i,j=1, j \neq i}^n  {\rm cov}_{n}(p_i,p_j) 
\simeq \frac{1}{\langle n \rangle^2} \sum_{i,j=1, j \neq i}^{\langle n \rangle} 
 {\rm cov}_{\langle n \rangle}(p_i,p_j). 
\label{dyn}
\end{eqnarray}

Now we come to the physics discussion.  The scaling (\ref{scaledyn}) imposes severe 
constraints on the physical nature of the covariance term. For instance, if all 
particles were correlated to each other, 
$\sum_{i,j=1, j \neq i}^n {\rm cov}_{\langle n \rangle}(p_i,p_j)$ would be proportional 
to the number of pairs, and $\sigma_{\rm dyn}$ would not depend on $\langle n \rangle$ at 
large multiplicities. A natural explanation of the scaling (\ref{scaledyn}) comes from the 
cluster model, depicted in Fig.~\ref{fig:clust}. The system is assumed to have $N_{\rm cl}$
clusters, each containing (on the average) $r$ particles. Below we keep $r= {\rm const.}$ 
for simplicity. The particles are correlated if and only if they belong to the 
same cluster, where the covariance per pair is $2\,{\rm cov}^\ast$. The number of 
correlated pairs within a cluster is $r(r-1)/2$. Some particles may be unclustered,
hence the ratio of clustered to all particles is 
$\langle N_{\rm cl}\rangle r /\langle n \rangle = \alpha$. If all particles are clustered
then $\alpha=1$. With these assumptions Eq.~(\ref{dyn}) becomes
\begin{eqnarray}
\sigma_{\rm dyn}^2 = \frac{\alpha(r-1)}{\langle n \rangle} {\rm cov}^\ast, 
\label{cluster}
\end{eqnarray}
which complies to the scaling (\ref{scaledyn}). An immediate conclusion here is that the ratio
$\alpha$ cannot depend on $\langle n \rangle$ (in the fiducial centrality range)
in order for the scaling to hold. 

The question now is whether we can use the above results to
 draw conclusions on effects of jets (minijets), which 
have been proposed as a possible explanation of the experimental data even at the considered 
soft momenta \cite{Adler:2003xq}. Jets, when fragmenting, lead to clusters in the 
momentum space. The resulting full covariance from jets
is then  $N_{\rm cl, jet} j(j-1) {\rm cov^j}/\langle n \rangle^2$, 
where $N_{\rm cl, jet}$ is the number of clusters originating from jets, $j$ is the average 
number of particles in the cluster, and $2\,{\rm cov^j}$ is the average covariance per pair.
The total number of particles produced from jets is $N_{\rm cl, jet} j$. 
On the other hand, the commonly accepted estimate of the dependence of $N_{\rm cl, jet} j$ on centrality 
is accounted for by the nuclear modification factor $R_{AA}$ multiplied by
the number of binary nucleon-nucleon collisions $N_{\rm bin}$. Since $R_{AA}$ depends on 
the ratio $\langle n \rangle / \langle n \rangle_{pp}$, where $\langle n \rangle_{pp}$ 
is the multiplicity in the proton-proton collisions, in a given $p_T$ bin one finds
\begin{eqnarray}
N_{\rm cl, jet}j \sim R_{AA} N_{\rm bin}= \frac{\langle n \rangle }{N_{\rm bin} 
\langle n \rangle_{pp} } N_{\rm bin} \sim \langle n \rangle, \label{jet}
\end{eqnarray} 
which complies to the scaling of Eq.~(\ref{cluster}).
We stress that this scaling follows just from the presence 
of clusters, and is insensitive to the nature of their physical origin as long as 
one imposes $N_{\rm cl} \sim \langle n \rangle$. 
In other words, as long as Eq.~(\ref{jet}) is used, 
the explanation of the observed data in terms of quenched jets agrees with the 
cluster picture. However, the explanation of the centrality dependence of the 
$p_T$ fluctuations in terms of jets based solely on Eq.~(\ref{jet}) is insufficient and not
conclusive: any mechanism leading to clusters would do.
Microscopic realistic estimates of the magnitude of ${\rm cov^j}$ and $j$ are necessary in that regard,
including the interplay of  jets and medium. For the current status of this program the user
is referred to \cite{Liu:2003jf,PHENIX:mitchell}.

Before continuing the analysis of the cluster model in a more quantitative manner 
we need to consider the effects of acceptance and detector efficiency. This is particularly
important in the event-by-event analysis, since the experiments select
particles { with very clearly identified tracks}, and thus the detector efficiency, denoted by $a$, is small. 
The number of observed particles is proportional to $a$, 
and the number of pairs contributing to the covariance is proportional to $a^2$. Thus Eq.~(\ref{cluster}) 
may be rewritten as $\sigma_{\rm dyn}^2 = \frac{r-1}{\langle n \rangle}_{\!\!\! \rm full}
{\rm cov}^\ast = a \frac{r-1}{\langle n \rangle}_{\!\!\! \rm obs} {\rm cov}^\ast$,
where "full" denotes all particles (that would be observed with 100\% efficiency), while "obs"
stands for the actually observed multiplicity of particles. Thus 
\begin{eqnarray}
{\rm cov}^\ast = \sigma_{\rm dyn}^2 \frac{\langle n \rangle_{\! \rm obs}}{a(r-1)}. 
\label{clust2}
\end{eqnarray}
Our estimate for $a$ in the PHENIX experiment is of the order of
10\%, which together with the numbers of Table \ref{tab:data} gives  
\begin{eqnarray}
{\rm cov}^\ast \simeq \frac{0.035~{\rm GeV}^2}{(r-1)}. 
\label{clust3}
\end{eqnarray}
In the considered problem the coefficient $0.035~{\rm GeV}^2$ is not a small
number when compared to  the natural
scale set by the variance $\sigma_p^2 \simeq 0.08~{\rm GeV}^2$. We recall that 
$\mid {\rm cov}^\ast \mid \le \sigma_p^2$. Comparing the numbers, we note that for 
$r=2$ the value of ${\rm cov}^\ast$ would assume almost a half of the maximum possible value. 
This is very unlikely, as argued in the dynamical estimates presented below, which give ${\rm cov}^\ast$ 
at most $0.01~{\rm GeV}^2$. Thus a natural explanation of the values in (\ref{clust3}) is to take 
a significantly larger value of $r$. Of course, the higher value, the easier it is to satisfy 
(\ref{clust3}) even with small values of ${\rm cov}^\ast$. We call this picture the ``lumped clusters'': 
lumps of matter move at some collective velocities, correlating the momenta of particles belonging 
to the same cluster, see Fig.~\ref{fig:clust}. 

The above estimates were based on the PHENIX data \cite{Adcox:2002pa}, 
however, very similar quantitative conclusions can be reached from the 
recently published STAR data \cite{Adams:2005ka}. We note that the measure 
$\langle \Delta p_i \Delta p_j \rangle$ used by STAR is just the 
estimator for $\sigma_{\rm dyn}^2$. Indeed, elementary steps lead to 
\begin{eqnarray}
\langle \Delta p_i \Delta p_j \rangle 
= \frac{N_{\rm event}-1}{N_{\rm event}} {\sigma}^2_M - 
\frac{1}{N_{\rm event}} \sum_{k=1}^{N_{\rm event}}  \frac{{\sigma}^2_p}{N_k}. \label{my}
\end{eqnarray}
Comparison to (\ref{central}) leads immediately for a large number of events to 
$\langle \Delta p_i \Delta p_j \rangle = \sigma_{\rm dyn}^2$. 
Now, taking the values of Table~I of Ref.~\cite{Adams:2005ka} and assuming $a=0.75$ we find 
$ {\rm cov}^\ast (r-1)=0.058, 0.043, 0.035, 0.014~{\rm GeV}^2$ for $\sqrt{s_{NN}}=200, 130, 62$ and 20 GeV,
respectively. The value at 130~GeV is close to the value (\ref{clust3}). Interestingly, we note a 
significant beam-energy dependence, with ${\rm cov}^\ast (r-1)$ increasing with $\sqrt{s_{NN}}$. 
This may be due to the increase of the covariance per correlated pair with the 
increasing energy, and/or an increase of the number of clustered particles. 

In the last part of this paper we present some dynamical estimates
of ${\rm cov}^\ast$ in thermal models. The first calculation concerns the role of resonances in 
$p_T$ correlations. Clearly, a resonance, such as the $\rho$ meson, decaying into daughter 
particles induces momentum correlations. We make a numerical calculation of this effect 
in the model of Ref.~\cite{Broniowski:2001we,Broniowski:2002nf}, using the
formula
\begin{eqnarray}
\hspace{-1cm} && {\rm cov}^\ast_{\rm res} 
= {\!\int\! d^3p \!\int {d^3p_1 \over E_{p_1}} \!\int {d^3p_2 \over E_{p_2}} 
\,\delta^{(4)}(p-p_1-p_2) C  {dN_R \over d^3p} 
\left(p_1^\perp\! -\! \langle p^\perp \rangle \right) 
\left(p_2^\perp\! - \!\langle p^\perp \rangle \right) 
\over
\!\int\! d^3p \int {d^3p_1 \over E_{p_1}} \int {d^3p_2 \over E_{p_2}} 
\, \delta^{(4)}(p-p_1-p_2) C  {dN_R \over d^3p} 
}, 
\label{thermalcov}
\end{eqnarray}
where $dN_R / d^3p$ is the resonance distribution in the momentum space (obtained
from the Cooper-Frye formula as described in Ref. \cite{Bozek:2003qi}), $p_1$ and $p_2$ are 
the momenta of the emitted particles, $E_p$ is the energy of a particle with momentum $p$,
and the function $C$ represents the experimental cuts. We note that from now on the 
letter $p$, depending on the context, denotes the four- or three-momentum. 
The results of our numerical study show that for the resonance mass between 
500~MeV and 1.2~GeV the covariance ${\rm cov}^\ast_{\rm res} $ varies between 
0.005~GeV$^2$ at low masses to $-0.015$~GeV$^2$ at high masses,  changing sign around 700-800~MeV, 
depending on the assumed experimental cuts.  Thus, cancellations between contributions of 
various resonances are possible; in fact, a full-fledged simulation with {\tt Therminator} 
\cite{Kisiel:2005hn} revealed a negligible contribution of resonances to the $p_T$ correlations. 
Of course, the ``lumpy'' feature of the expansion was not implemented in the calculation. 
Details of this study will be presented elsewhere.

The second model of particle correlations assumes that the particle emission
at the lowest scales occurs from local thermalized sources. 
Each element of the fluid moves with its collective velocity and emits
particles  with locally thermalized spectra. This picture was put forward as a
mechanism creating correlations in the  charge balance function 
\cite{Bozek:2003qi,Cheng:2004zy}
resulting from   charge conservation within the local source. 
Correlations between 
 particles emitted from the same cluster come from the fact that those
particles are emitted from a source with the same collective transverse
velocity. The average number of particles $r$ 
originating from such a local source determines the strength of the surviving
dynamical fluctuation in the whole event, as discussed above. The covariance
between particles $i$ and $j$ emitted from a cluster moving with a velocity $u$ is
\begin{equation}
\hspace{-0.75cm} {\rm cov}^\ast(i,j)=\frac{ \int d\Sigma_\mu u^\mu 
\int d^3p_1  (p^\perp_1-\langle p^\perp \rangle )
f_i^u(p_1) \int d^3p_2 (p^\perp_2- \langle p^\perp \rangle ) f_j^u(p_2)}
 { \int d\Sigma_\mu u^\mu 
\int d^3p_1 f_i^u(p_1) \int d^3p_2
f_j^u(p_2)}, 
\end{equation}
where $f_i^u(p)=(\exp(p \cdot u /T)\pm 1)^{-1}$ is the boosted thermal
distribution and $d\Sigma_\mu $ denotes integration over the freeze-out hypersurface. 
The result turns out to depend strongly on the temperature. 
Considering the emission of correlated pion pairs 
 one gets ${\rm cov}^\ast(\pi,\pi)=0.0034~{\rm GeV}^2$ 
for  freeze-out parameters
corresponding to the single freeze-out model
\cite{Broniowski:2001we}($T=165$~MeV, 
average flow velocity $0.5c$)  and
${\rm cov}^\ast(\pi,\pi)=0.01~{\rm GeV}^2$ for parameters corresponding to a late kinetic
freeze-out  ($T=100$~MeV,
average flow velocity $0.6c$). For realistic values of  thermal freeze-out
parameters the experimentally estimated value of the covariance cannot be
accounted for, unless the number of charged particles belonging to the same cluster
is at least $4-10$. 

In conclusion, we have found that in the fiducial centrality range 
the scaling of the $\sigma_{\rm dyn}^2$ for the $p_T$ correlations 
with inverse particle multiplicity indicates the cluster nature of the 
system formed in relativistic heavy-ion collisions. The clusters may a priori originate from
very different physics: jets, droplets of fluid formed in the explosive scenario of the collision, 
or other mechanisms leading to multiparticle correlations. A larger number of particles within a
cluster helps to obtain the large measured value of $\sigma_{\rm dyn}^2$.

\end{document}